\documentstyle[11pt,aaspp4]{article}
\newcommand{\ts}{\thinspace}

\begin{document}

\title{NEAR-INFRARED SPECTROSCOPY OF THE HIGH REDSHIFT QUASAR
       S4{\ts}0636+68 AT $z$=3.2}

\author{\sc Takashi Murayama\altaffilmark{1}
        and Yoshiaki Taniguchi\altaffilmark{1}}
\affil{Astronomical Institute, Tohoku University,
                Aoba, Sendai 980-8578, Japan}
\affil{Electronic mail: murayama@astr.tohoku.ac.jp,
                            tani@astr.tohoku.ac.jp}

\author{\sc Aaron S.~Evans\altaffilmark{1}}
\affil{California Institute of Technology, 105-24,
       Pasadena, CA 91125}
\affil{Electronic mail: ase@astro.caltech.edu}

\author{\sc D.~B.~Sanders}
\affil{Institute for Astronomy, University of Hawaii,
      2680 Woodlawn Drive, Honolulu, HI 96822}
\affil{Electronic mail: sanders@ifa.hawaii.edu}

\author{\sc Youichi Ohyama\altaffilmark{1}}
\affil{Astronomical Institute, Tohoku University,
                Aoba, Sendai 980-8578, Japan}
\affil{Electronic mail: ohyama@astr.tohoku.ac.jp}


\author{\sc Kimiaki Kawara and Nobuo Arimoto}
\affil{Institute of Astronomy, The University of Tokyo,
                 2-21-1 Osawa, Mitaka, Tokyo 181-8588, Japan}
\affil{Electronic mail: kkawara@mtk.ioa.s.u-tokyo.ac.jp,
                        arimoto@mtk.ioa.s.u-tokyo.ac.jp}

\affil{ }
\affil{To appear in the Astronomical Journal}

\altaffiltext{1}{Visiting Astronomer of the University of Hawaii
                 2.2 meter telescope.} 

\authoremail{murayama@astr.tohoku.ac.jp}

\begin{abstract}
We present near-infrared (observed frame) spectra of the high-redshift 
quasar S4{\ts}0636+68 at $z=3.2$ which was previously thought to be one of 
a group of ``strong'' \ion{Fe}{2} emitters (i.e., $F(\mbox{\ion{Fe}{2}}{\ts}
\lambda\lambda\mbox{4434--4684})/F({\rm H}\beta) > 1$).
Our {\it K}-band spectrum clearly shows emission lines of H$\beta$
and [\ion{O}{3}]{\ts}$\lambda\lambda$4959,{\ts}5007 as well as optical 
\ion{Fe}{2} emission.  Our computed value of $F(\mbox{\ion{Fe}{2} }
\lambda\lambda\mbox{4434--4684})/F({\rm H}\beta) \simeq 0.8$ for 
S4{\ts}0636+68 is less than previously thought, and in fact is comparable 
to values found for radio-loud, flat-spectrum, low-$z$ quasars.
Therefore S4{\ts}0636+68 appears not to be a strong optical \ion{Fe}{2} 
emitter.  Although more than half (5/8) of the high-$z$ quasars 
observed to date are still classified as strong optical \ion{Fe}{2} 
emitters, their \ion{Fe}{2}/H$\beta$ ratios, for the most part, 
follow the same trend as that of low-$z$ quasars, i.e., an 
anticorrelation in $EW$(\ion{Fe}{2})/$EW$(H$\beta$) versus 
$EW$([\ion{O}{3}])/$EW$(H$\beta$), with radio-loud quasars having a  
mean value of $EW$(\ion{Fe}{2})/$EW$(H$\beta$) approximately half 
that of radio-quiet quasars at comparable values of 
$EW$([\ion{O}{3}])/$EW$(H$\beta$).
\end{abstract}


\section{INTRODUCTION}

Since optical \ion{Fe}{2} emission\footnote{%
The \ion{Fe}{2} emission feature actually extends 
from the near-UV into the red optical region of the spectrum. 
However, following previous convention, we use the term 
``optical \ion{Fe}{2} emission'' in this paper to mean the \ion{Fe}{2} 
emission near H$\beta$ (i.e., \ion{Fe}{2}{\ts}$\lambda\lambda$4434--4684).}
is often one of the prominent features
in the spectra of Type 1 active galactic nuclei (AGN), it is perhaps not surprising that 
several  observational and theoretical studies have been made 
to explain the strength of this feature in quasars
(e.g. Phillips \markcite{Phillips77}1977, \markcite{Phillips78}1978;
\markcite{Kwan81}Kwan \& Krolik 1981;
\markcite{Netzer83}Netzer \& Wills 1983;
\markcite{Wills85}Wills {\it et al.}\ 1985;
\markcite{Collin-Souffrin88}Collin-Souffrin {\it et al.}\ 1988;
\markcite{Zheng90}Zheng \& O'Brien 1990;
\markcite{Joly91}Joly 1991;
\markcite{Boroson92}Boroson \& Green 1992;
\markcite{Lipari93}L\'{\i}pari {\it et al.}\  1993;
\markcite{Wang96a}Wang {\it et al.}\ 1996a).
Although it is known that the strength of the optical \ion{Fe}{2}  emission shows 
an anticorrelation with the strength of [\ion{O}{3}] emission (Boroson \& Green 1992),
the physical properties of the \ion{Fe}{2}  emitting region are not yet
fully understood. 

Recent near-infrared (NIR) spectroscopic studies by 
\markcite{Hill93}Hill {\it et al.}\ (1993) and
\markcite{Elston94}Elston {\it et al.}\ (1994; hereafter ETH) 
suggest that unusually strong optical \ion{Fe}{2} emitters may be
common in the high-$z$ universe ($2 < z < 3.4$).
Though it is known that some low-$z$
far-infrared (FIR) selected  AGN
($L_{\rm FIR} \gtrsim 10^{11}$ $L_{\sun}$)
show strong
\ion{Fe}{2} emission in their optical spectra
\markcite{Lipari93}(cf.\ L\'{\i}pari {\it et al.}\ 1993),
such extreme \ion{Fe}{2} emitters
appear to be rare in the low-$z$ universe.
Recently, we obtained NIR spectra of two radio-loud, flat-spectrum, high-$z$ quasars
(B 1422+231 at $z=3.6$ and PKS 1937$-$101 at $z=3.8$) and found that
their flux ratios of
$F(\mbox{\ion{Fe}{2} }{\ts}\lambda\lambda\mbox{4434--4684})/F({\rm H}\beta)$ 
are much less than those of the other high-$z$ quasars 
(\markcite{Kawara96}Kawara {\it et al}.\ 1996;
Taniguchi {\it et al.}\ \markcite{Taniguchi96}1996,
\markcite{Taniguchi97}1997), and in fact are similar to those of 
radio-loud, flat-spectrum, low-$z$ quasars with normal optical \ion{Fe}{2}
emission.  These new observations suggest that high-$z$ quasars may exhibit a range 
of values of $F(\mbox{\ion{Fe}{2} }{\ts}\lambda\lambda\mbox{4434--4684})/F({\rm H}\beta)$ 
similar to what has been observed for low-$z$ quasars.


If the strong \ion{Fe}{2} emission could be attributed to the
overabundance of iron, host galaxies of the high-$z$ quasars
with strong \ion{Fe}{2} emission would form at $z \sim${\ts}10
because it is usually considered to be the case that the bulk of the iron arises
from Type Ia supernovae which occur $\sim${\ts}1--2{\ts}Gyr after 
the first major epoch of star formation (e.g., \markcite{Hamann93}Hamann \& Ferland 
1993, 
\markcite{Yoshii96}Yoshii {\it et al}.\ 1996).
It is therefore important to investigate the chemical properties 
of high-$z$ quasars systematically.

In this paper we present new NIR spectroscopy of S4{\ts}0636+68,
a flat-spectrum radio-loud quasar at $z=3.2$, which 
is reported in ETH as a very strong iron emitter.
Based on our new measurements, we discuss whether
the fraction of high-$z$ quasars with  
strong optical \ion{Fe}{2} emission 
is substantially higher than that of low-$z$ quasars.

\section{OBSERVATIONS AND DATA REDUCTION}

We observed S4{\ts}0636+68 on 1996 April 7 (UT) using the $K$-band
spectrograph (KSPEC; \markcite{Hodapp94}Hodapp {\it et al}.\ 1994)
at the Cassegrain focus (f/31) of the University of Hawaii (UH) 2.2 m
telescope in combination with the UH tip-tilt system
(\markcite{Jim97}Jim {\it et al}.\ 1997).
The cross-dispersed echelle design of KSPEC
provided simultaneous coverage of the entire 
1--2.5 \micron{} wavelength region.
The projected pixel size of the HAWAII 1024 $\times$ 1024 array
was 0\farcs{}167 along the slit and $\simeq$ 5.6 \AA{} at
2 \micron{} along the dispersion direction.  We used a 0\farcs{}96 
wide slit oriented East-West and centered on the intensity peak
of the object.  Twenty exposures, each of 180 sec 
integration under photometric conditions, were obtained by
shifting the position of the object along the slit
at intervals of 5\arcsec{} between each integration.
The total integration time was 3600 sec.
An A-type standard star, HD 106965
(\markcite{Elias82}Elias {\it et al}.\ 1982),
was observed for flux calibration.
Another A-type star, HD 136754
(\markcite{Elias82}Elias {\it et al}.\ 1982), was also observed
before and after observing
S4{\ts}0636+68 in order to correct for atmospheric absorption.
Spectra of an incandescent lamp and an argon lamp were taken for
flat-fielding and wavelength calibration, respectively.
Typical widths of the spatial profiles of the
standard star spectra were $\sim${\ts}0\farcs{}5 (FWHM) throughout the night.

Data reduction was performed with IRAF\footnote{%
Image Reduction and Analysis Facility (IRAF) is
distributed by the National Optical Astronomy Observatories,
which are operated by the Association of Universities for Research
in Astronomy, Inc., under cooperative agreement with the National
Science Foundation.} using standard procedures as outlined in 
\markcite{Hora96}Hora \& Hodapp (1996).
Sky and dark counts were removed by subtracting the average of
the preceding and following exposures, and then the resulting
frame was divided by the normalized dome flat.
The target quasar was not bright enough to trace
its position in each frame with sufficient accuracy. 
Therefore, at first, we fit the spectral positions of the standard star
spectra with a third-order polynomial function
which properly traces the echelle spectrum.
These fitting results were then applied to the quasar spectra.
Using this procedure, we extracted the quasar spectra with an aperture
of 3\arcsec{} which was determined to be the typical width
where the flux was $\sim${\ts}10{\ts}\% of
the peak flux along the spatial profile of the standard star.
In order to subtract the residual sky emission, we used the data just adjacent
to the 3\arcsec{} aperture.
The wavelength scale of each extracted spectrum was calibrated to an 
accuracy of 18 km s${}^{-1}$ at 1.1 \micron, 20 km s${}^{-1}$ at
1.6 \micron, and 19 km s${}^{-1}$ at 2.0 \micron, respectively,
based on both the argon emission lines of the calibration lamp and
on the telluric OH emission lines.
The spectral resolutions (FWHM) measured from the argon
lamp spectra were $\simeq$ 500 km s${}^{-1}$ at 1.1 \micron, $\simeq$
450 km s${}^{-1}$ at 1.6 \micron, and $\simeq$ 500 km s${}^{-1}$ at 2.0
\micron.
The spectra were finally median combined in each band.
Atmospheric absorption features were removed
using the spectra of the A-type star HD 136754
because A-type stars are best suited for correcting for atmospheric absorption
features.
However, since A-type stars inherently have hydrogen recombination
absorption lines (e.g., the Brackett series in $H$ and $K$ bands
and the Paschen series in $I$ and $J$ bands),
we removed these features before the correction
using Voigt profile fitting.
In order to check whether this procedure worked 
we also applied the same atmospheric correction for the spectra of M-type stars
whose data were obtained on the same night.
Comparing our corrected spectra of the M-type stars with their 
published spectra (Lan\c{c}on \& Rocca-Volmerange 1992),
we found that our correction procedure works appropriately.
Finally, in order to calibrate the flux scale, we used 
the spectrum of the standard star HD 106965 (A2, $K$=7.315)
divided by a 9000 K blackbody spectrum, which fits
the $J\!H\!K\!L$ magnitude of the standard
\markcite{Elias82} (Elias {\it et al}.\ 1982)
with only 1.2{\ts}\% deviation.
Photometric errors were determined to be $<${\ts}10{\ts}\%
over all observed wavelengths.

\section{RESULTS AND DISCUSSION}

\subsection{The Rest-Frame UV and Optical Spectra of S4{\ts}0636+68}
Figure \ref{fig-1} shows the spectra of S4{\ts}0636+68 (solid line)
in the $I\!H\!K$ bands 
together with the Large Bright Quasar Survey (LBQS) composite
spectrum (dashed line; \markcite{Francis91}
Francis {\it et al}.\ 1991) shifted to $z=3.2$. The atmospheric
transmission of Mauna Kea is shown in the upper panel.
The spectrum clearly shows H$\beta$ at 2.04 \micron{} and a broad ``bump''
of \ion{Fe}{2} emission between 2.15{\ts}\micron{} and 2.23{\ts}\micron{}.
The spike feature in our spectrum marked by `X' is caused by
residual atmospheric absorption.
Although ETH did not find evidence
for [\ion{O}{3}]{\ts}$\lambda\lambda$4959,{\ts}5007
emission lines, our $K$-band spectrum shows their presence
(which will be discussed later).
No prominent emission lines were found in the
$I$-, $J$-, and $H$-band regions of the spectrum.
\ion{Mg}{2}{\ts}$\lambda$2798 would be expected to appear at $\sim$ 1.18
\micron{} but the $J$-band spectrum (not shown in Figure \ref{fig-1})
was too noisy to be able to study \ion{Mg}{2}.
Since the efficiency of KSPEC in the $J$-band is not high,
we do not use the $J$-band data in this paper.

\placefigure{fig-1}

In Figure 2, we compare our results with previous optical-NIR
spectroscopic studies of S4{\ts}0636+68
(\markcite{Sargent89}Sargent {\it et al}.\ 1989;
\markcite{Bechtold94}Bechtold {\it et al}.\ 1994; ETH).
Their basic data are summarized in Table \ref{tbl-1}.
Our $K$-band spectrum is twice as bright as that of ETH.
On the other hand, 
our $I$-band spectrum is 40{\ts}\% fainter than
that of \markcite{Bechtold94}Bechtold {\it et al}.\ (1994).
These flux discrepancies may be due to 
possible time variation
inherent in the object or to calibration errors in the 
absolute photometry.
However, since the two optical spectra taken at different observing dates
over two years are quite consistent with each other
(\markcite{Sargent89}Sargent {\it et al}.\ 1989;
\markcite{Bechtold94}Bechtold {\it et al}.\ 1994),
it seems unlikely that this quasar is highly variable.
Therefore, we consider the possibility that the discrepancy may be mostly due to
calibration errors.
First, we note that our $I\!H\!K$ spectra were taken simultaneously
and thus there is no internal calibration error in our spectra.
Second, the optical spectra of
\markcite{Sargent89}Sargent {\it et al}.\ (1989) and
\markcite{Bechtold94}Bechtold {\it et al}.\ (1994) show
good agreement with each other and thus
their photometric calibration seems reliable.
Further, the optical power-law slope,
$\alpha=-0.68$ ($f_\nu \propto \nu^\alpha$),
given in \markcite{Sargent89}Sargent {\it et al}.\ (1989)
can be consistently extrapolated onto our $K$-band 
spectrum; the spectral index using the emission-free regions
in both the optical spectrum
(\markcite{Sargent89}Sargent {\it et al}.\ 1989) and
our $K$-band spectrum
(1330--1380 \AA, 1430--1460 \AA, and 5400--5850 \AA{}
in the rest frame) is estimated to be $\alpha=-0.69$.
Since the seeing during our observations was $\simeq 0\farcs{}5$ (FWHM) 
and our slit size was 1\arcsec{}, we think that we have detected 
nearly all of the light from the quasar.
Though the details of the observing conditions and slit size
are not given in ETH (see Table \ref{tbl-1}),
seeing conditions on Mauna Kea are often better than at KPNO, 
judging from our experience at KPNO
(see \markcite{Kawara96}Kawara {\it et al}.\ 1996;
\markcite{Kawara97}Taniguchi {\it et al}.\ 1997), and, therefore 
we expect that our new measurement is more reliable.

\begin{table}
\dummytable\label{tbl-1}
\end{table}
\placetable{tbl-1}
\placefigure{fig-2}

\subsection{The Rest-frame Optical Emission-line Properties of S4{\ts}0636+68}

The main aim of our current observations is to provide a more accurate 
measure of the optical \ion{Fe}{2}/H$\beta$ ratio in S4{\ts}0636+68. 
Figure \ref{fig-3} compares our result with that of the ETH.
(Note that the flux of ETH spectrum is scaled
by a factor of two for proper comparison.)
Center positions of H$\beta$, [\ion{O}{3}]{\ts}$\lambda\lambda$4959,{\ts}5007,
and the \ion{Fe}{2} multiplet (42) at a redshift  $z=3.2$
are marked in Figure \ref{fig-3}.
The peak positions of H$\beta$, [\ion{O}{3}]{\ts}$\lambda\lambda$4959,{\ts}5007,
and \ion{Fe}{2}{\ts}$\lambda$5169 (one of \ion{Fe}{2} multiplet 42 lines), coincide
between the two spectra.
The $K$-band spectrum of ETH appears to be dominated by very strong
\ion{Fe}{2} emission with weak, or nondetected [\ion{O}{3}]{\ts}$\lambda\lambda$4959,{\ts}5007. 
ETH actually stated that [\ion{O}{3}]{\ts}$\lambda\lambda$4959,{\ts}5007
was not detected, 
although they noted that their spectrum had small bumps of low
significance at the position of the [\ion{O}{3}] lines.
On the other hand, our $K$-band spectrum clearly shows
emission peaks which can be identified
with [\ion{O}{3}]{\ts}$\lambda\lambda$4959,{\ts}5007.

\placefigure{fig-3}

To measure the \ion{Fe}{2} fluxes in our spectrum, we fit
emission-line features simultaneously with a least-squares algorithm.
Such fitting results depend on the adopted continuum spectrum.
As shown in Boroson \& Green (1992), a local linear continuum is usually
adopted to fit H$\beta$, [\ion{O}{3}]{\ts}$\lambda\lambda$4959,{\ts}5007, 
and the \ion{Fe}{2} features. 
However, we have already
obtained a global power-law continuum using the
rest-frame UV and optical spectra
as shown in Figure 2. Therefore, we performed spectral
fitting for two cases;
1) local linear continuum and 2) global power-law continuum.
In the fitting procedure, we assumed  
$F(\mbox{[\ion{O}{3}]}{\ts}\lambda{\rm 5007})/F(\mbox{[\ion{O}{3}]}{\ts}
\lambda {\rm 4959})$ = 2.97 (\markcite{Osterbrock89}Osterbrock 1989). 
We also assumed that the emission line profiles of
H$\beta$ and the [\ion{O}{3}]{\ts}$\lambda\lambda$4959,{\ts}5007 doublet
are Gaussian. As for the optical \ion{Fe}{2} emission features, we used an
\ion{Fe}{2} spectrum of a low-$z$ BAL quasar, PG 0043+039
(\markcite{Turnshek94}Turnshek {\it et al}.\ 1994), as our \ion{Fe}{2} template.
All the emission lines are assumed to have the same redshift.
Since it is known that high-ionization broad lines (e.g.,
\ion{C}{4}{\ts}$\lambda$1549)
are often blueshifted with respect to low-ionization lines (Gaskell 1982; Wilkes 1984;
Carswell {\it et al}.\ 1991; Nishihara {\it et al}.\ 1997
and references therein),
we use only low ionization lines in our analysis.

The fitting results are presented in Figure \ref{fig-4} and Table \ref{tbl-2}
for each of the two assumed continua.  The difference in the line flux ratios 
between the two cases is less than the measurement errors.
Although we do not know which continuum case is more realistic,
we adopt the results using the linear continuum fit for further  discussion
in order to compare our results with those of 
Boroson \& Green (1992) and Hill {\it et al}. (1993)
since they also adopted a local linear continuum.

In order to examine whether or not the detection of
the [\ion{O}{3}] lines are real in our spectrum,
we compare our fit including the [\ion{O}{3}] doublet with a fit 
excluding the [\ion{O}{3}] doublet, where the local linear continuum 
has been adopted in both fits.
A F-statistics test indicates that the fit with [\ion{O}{3}] is
improved over the 4900--5050 \AA{} region from the fit 
excluding [\ion{O}{3}] at a significance level of  99.8{\ts}\%.
Therefore, we conclude that the ``bumps'' at the [\ion{O}{3}] positions
are really the [\ion{O}{3}]{\ts}$\lambda\lambda$4959,{\ts}5007 doublet rather than
\ion{Fe}{2}{\ts}$\lambda\lambda$4924,{\ts}5018 of the 42 multiplet.
We obtained an average redshift $z=3.200 \pm 0.002$.

\begin{table}
\dummytable\label{tbl-2}
\end{table}
\placetable{tbl-2}
\placefigure{fig-4}

Our fit assuming a local linear continuum gives the flux ratio $F(\mbox{\ion{Fe}{2} }
\lambda 5169)/F({\rm H}\beta)=0.28$.
This value is significantly smaller than the value of 0.45 that 
we estimate from the published spectrum of ETH. 
Although we do not understand this difference,
it could reasonably be explained by uncertainties in the continuum 
fits between the two observations.
However, we cannot rule out the possibility of time variation.
In fact, such a time variation of \ion{Fe}{2} is reported for
the nearby type 1 Seyfert galaxy NGC 5548
(\markcite{Wamsteker90}Wamsteker {\it et al}.\ 1990;
\markcite{Maoz93}Maoz {\it et al}.\ 1993;
\markcite{Sergeev97}Sergeev {\it et al}.\ 1997).
Thus, monitoring of S4{\ts}0636+68 may be needed in the future.

The flux ratio, $F(\mbox{\ion{Fe}{2} }
\lambda\lambda\mbox{3500--6000})/F({\rm H}\beta)$,
for  S4{\ts}0636+68 is $3.5\pm1.1$ (Table \ref{tbl-2}).
This value is greater than the mean value of $1.63 \pm 0.88$ for
the six low-$z$ quasars studied by
\markcite{Wills85}Wills {\it et al}.\ (1985)
and the value of 2.9 for 3C 273
which is the strongest optical \ion{Fe}{2} quasar
in the sample of \markcite{Wills85}Wills {\it et al}.\ (1985).
However, $F(\mbox{\ion{Fe}{2}}{\ts}\lambda\lambda\mbox{4434--4684})
/F({\rm H}\beta)$
for S4{\ts}0636+68 is $0.83 \pm 0.26$ which is only half of the average
value of $1.77 \pm 0.17$ for the four high-$z$ quasars studied
by \markcite{Hill93}Hill {\it et al}.\ (1993).
\markcite{Lipari93}L\'{\i}pari {\it et al}.\ (1993) defined 
quasars with $F(\mbox{\ion{Fe}{2}}{\ts}\lambda\lambda\mbox{4434--4684})
/F({\rm H}\beta) \gtrsim 1$
as ``strong'' iron emitters.
According to this criterion, we conclude that S4{\ts}0680+68
is not a strong \ion{Fe}{2} emitter, contrary to the conclusion of ETH.

\subsection{Statistical Properties of High-$z$ Quasars
            vs.\ Low-$z$ Quasars}

In order to assess the significance of our new result for the ratio 
$F(\mbox{\ion{Fe}{2}}{\ts}\lambda\lambda\mbox{4434--4684})
/F({\rm H}\beta)$ in S4{\ts}0680+68 we first compare the rest-frame optical 
emission line properties of low-$z$ and high-$z$ quasars.
Figure \ref{fig-5} shows the relationship of equivalent width (EW) ratios
between $EW(\mbox{[\ion{O}{3}]}{\ts}\lambda4959+\lambda5007)
/ EW({\rm H}\beta)$
and $EW(\mbox{\ion{Fe}{2} }{\ts}\lambda\lambda\mbox{4434--4684})
/ EW({\rm H}\beta)$
for low-$z$ and high-$z$ quasars compiled from the literature
(\markcite{Boroson92}Boroson \& Green 1992;
\markcite{Hill93}Hill {\it et al}.\ 1993; ETH;
\markcite{Kawara96}Kawara {\it et al}.\ 1996;
\markcite{Taniguchi97}Taniguchi {\it et al}.\ 1997).
There is a distinct anticorrelation for the low-$z$ quasars
as noted before (cf.\ \markcite{Boroson92}Boroson \& Green 1992)
although the reason for the anticorrelation between 
[\ion{O}{3}]{\ts}$\lambda\lambda$4959,{\ts}5007 and
optical \ion{Fe}{2}{\ts}$\lambda\lambda\mbox{4434--4684}$ is still unknown.
The low-$z$ radio-loud quasars tend to have  small ratios both in
$EW(\mbox{[\ion{O}{3}]}{\ts}\lambda4959+\lambda5007)
/EW({\rm H}\beta)$
and $EW(\mbox{\ion{Fe}{2}}{\ts}\lambda\lambda\mbox{4434--4684})
/EW({\rm H}\beta)$.
Five of the eight high-$z$ quasars show strong \ion{Fe}{2}
(i.e., \ion{Fe}{2}/H$\beta > 1$) emission. The remaining three high-$z$ quasars, 
which have \ion{Fe}{2}/H$\beta < 1$, are all radio-loud and 
lie within the locus of values 
traced by low-$z$ radio-loud quasars in Figure 5\    
\footnote{We note that a radio-loud high-$z$ ($z=2.09$) quasar 
1331+170 also appears to lie within the locus of 
values found for the low-$z$ radio-loud quasars (i.e., 1331+170 appears to 
have ``quite weak'' optical \ion{Fe}{2} emission and
$EW(\mbox{[\ion{O}{3}]}{\ts}\lambda4959+\lambda5007)/EW({\rm H}\beta) 
\sim{\ts}0.7$ \ (Carswell {\it et al.} \ 1991).}.
Also, the three radio-quiet 
quasars among the five high-$z$ quasars with strong \ion{Fe}{2} emission 
appear to lie within the upper envelope of values observed for low-$z$ 
radio-quiet quasars.
In summary, although over half (5/8) of the high-$z$ quasars appear to be
by definition strong \ion{Fe}{2} emitters, all but two
(the radio-loud \ion{Fe}{2} quasars S5{\ts}0014+81 and B2{\ts}1225+317: ETH; 
\markcite{Hill93}Hill {\it et al}.\ 1993) 
of the high-$z$ quasars follow a similar trend as that shown by the low-$z$
quasars, i.e.\ an anticorrelation of
$EW(\mbox{\ion{Fe}{2}}{\ts}\lambda\lambda\mbox{4434--4684})
/EW({\rm H}\beta)$ versus $EW(\mbox{[\ion{O}{3}]}{\ts}\lambda4959+\lambda5007)
/EW({\rm H}\beta)$, with radio-loud quasars having on average smaller values of
$EW(\mbox{\ion{Fe}{2}}{\ts}\lambda\lambda\mbox{4434--4684})
/EW({\rm H}\beta)$ than radio-quiet quasars at any given value of
$EW(\mbox{[\ion{O}{3}]}{\ts}\lambda4959+\lambda5007)
/EW({\rm H}\beta)$.


\placefigure{fig-5}

Recently Wang {\it et al}.\ (\markcite{Wang96b}1996b)
studied the relation between optical \ion{Fe}{2}
strength and properties of the UV spectra for
53 low-$z$ ($z \lesssim 0.2$) quasars and found that
there is a significant anticorrelation between the equivalent 
widths of optical \ion{Fe}{2}{\ts}{\ts}$\lambda\lambda\mbox{4434--4684}$ and 
\ion{C}{4}{\ts}$\lambda$1549.  We examine whether the high-$z$ quasars
follow this anticorrelation (Table \ref{tbl-3} and Figure \ref{fig-6}).
It is perhaps expected that the high-$z$ quasars would have smaller 
{\it EW}(\ion{C}{4}{\ts}$\lambda$1549)
than low-$z$ quasars simply because of the known anticorrelation between
{\it EW}(\ion{C}{4}{\ts}$\lambda$1549) and UV continuum luminosity 
(Baldwin effect: \markcite{Baldwin77}Baldwin 1977;
\markcite{Baldwin78}Baldwin {\it et al}.\ 1978).  Not as evident 
perhaps is that, except for 0933+733, the 
{\it EW}(\ion{Fe}{2}{\ts}$\lambda\lambda$4434--4684), appears to show 
the same range of values as do the low-$z$ quasars at comparable 
low values of {\it EW}(\ion{C}{4}{\ts}$\lambda$1549).
However, five of the remaining seven high-$z$ quasars 
(B2{\ts}1225+317, 1246$-$057, S4{\ts}0636+68, B{\ts}1422+231,
and PKS{\ts}1937$-$101) have smaller 
{\it EW}(\ion{Fe}{2}{\ts}$\lambda\lambda$4434--4684)
than any of the low-$z$ quasars with comparable 
{\it EW}(\ion{C}{4}{\ts}$\lambda$1549) 
(see the lower-left region of the diagram in Figure 6), 
thus, adding the high-$z$ sample to the low-$z$ sample appears to 
decrease somewhat the significance of the anticorrelation between 
{\it EW}(\ion{Fe}{2}{\ts}$\lambda\lambda$4434--4684) versus 
{\it EW}(\ion{C}{4}{\ts}$\lambda$1549), 
(although it is possible that not having a less luminous high-$z$ 
sample may cause a selection effect).  We thus consider it possible 
that the \ion{Fe}{2}{\ts}$\lambda\lambda$4434--4684 emitting 
region may not have a physical link directly with the 
\ion{C}{4}{\ts}$\lambda$1549 emitting region.  However, it seems clear 
that there is no object with large EWs in both 
\ion{Fe}{2}{\ts}$\lambda\lambda$4434--4684 and 
\ion{C}{4}{\ts}$\lambda$1549, and furthermore, the upper bound of 
{\it EW}(\ion{Fe}{2}{\ts}$\lambda\lambda$4434--4684) still decreases with increasing
{\it EW}(\ion{C}{4}{\ts}$\lambda$1549) for the combined high-$z$ 
and low-$z$ samples.  Hence, it is suggested that there may 
still be an indirect relation between the 
\ion{Fe}{2}{\ts}$\lambda\lambda$4434--4684 and 
\ion{C}{4}{\ts}$\lambda$1549 regions.

\begin{table}
\dummytable\label{tbl-3}
\end{table}
\placetable{tbl-3}
\placefigure{fig-6}

In summary,
the relations among the emission-line properties
shown in Figures \ref{fig-5} and \ref{fig-6} appear to be 
valid for both the low-$z$ and high-$z$ quasars with only a few exceptions.
This implies that the emission mechanism and
the physical properties of the emission-line region in high-$z$ quasars
may not be significantly different from those in low-$z$ quasars.

\acknowledgments

We are very grateful to the staff of the UH 2.2 m telescope.
In particular, we would like to thank Klaus Hodapp for his encouragement
and Andrew Pickles for his
technical support and assistance with the observations.
This work was financially supported in part by Grants-in Aid for 
Scientific Research (Nos.\ 07044054 and 09640311)
from the Japanese Ministry
of Education, Science, Sports, and Culture and by the Foundation for
Promotion of Astronomy, Japan.
TM thanks the support of Research Fellowships from the Japan
Society for the Promotion of Science for Young Scientists.
All the figures in this paper were prepared with GP,
the graph plotting tool developed by Keiichi Edamatsu.
This research has made use of the NASA/IPAC Extragalactic Database
(NED) and the NASA Astrophysics Data System Abstract Service.

\clearpage

\figcaption[Murayama.fig1.ps]{%
Comparison of the observed-frame $I\!H\!K$ spectra of S4{\ts}0636+68 
(solid line) and 
the LBQS composite spectrum of
Francis {\it et al}.\ (1991; dashed line).
The upper panel shows the atmospheric transmission for Mauna Kea
produced using the program IRTRANS4.
These data were obtained from the UKIRT
worldwide web pages (http://www.jach.hawaii.edu/UKIRT/home.html).
\label{fig-1}
}

\figcaption[Murayama.fig2.ps]{%
The combined optical and near-infrared spectra of S4{\ts}0636+68 from the literature
plus our $I\!H\!K$-band spectra.
The best power-law fit to the optical--infrared continuum
is indicated by the solid line.
The dotted line is the  LBQS composite spectrum
(Francis {\it et al}.\ 1991).
\label{fig-2}}

\figcaption[Murayama.fig3.ps]{%
The solid line shows our observed-frame $K$-band spectrum of S4{\ts}0636+68 
and that of Elston {\it et al}.\ (1994) (dashed line).
The spectrum of Elston {\it et al}.\ (1994) is scaled by a factor
of two for comparison.
The spike feature marked by `X' is caused by residual atmospheric
absorption.
The center positions of H$\beta$, [O{\sc{} iii}]{\ts}$\lambda\lambda$4959,{\ts}5007, 
 and the Fe{\sc{}~ii} multiplet 42 are indicated by marks. 
A redshift of $z=3.2$ is assumed.
The upper panel shows the atmospheric transmission.
\label{fig-3}}

\figcaption[Murayama.fig4.ps]{%
The profile fitting of the $K$-band spectrum.
The synthesized spectrum is over-plotted by the dashed line
on the original spectrum (solid line).
Note that the flux is multiplied by $(1+z)$ for deredshifting.
A linear continuum and each emission-line component are also shown.
\label{fig-4}}

\figcaption[Murayama.fig5.ps]{%
Diagram between equivalent width ratios of
{\it EW}([O{\sc{} iii}]{\ts}$\lambda$5007+$\lambda$4959)/{\it EW}(H$\beta$) 
and {\it EW}(Fe{\sc{}~ii}{\ts}$\lambda\lambda$4434--4684)/{\it EW}(H$\beta$)
for low-$z$ (small symbols)
and high-$z$ quasars (large symbols).
Sample data are compiled from
Boroson \& Green (1992),
Hill {\it et al}.\ (1993),
Elston {\it et al}.\ (1994),
Kawara {\it et al}.\ (1996),
Taniguchi {\it et al}.\ (1997).
Radio-quiet quasars, flat-spectrum radio-loud quasars, and
steep-spectrum radio-loud quasars
are shown by open circles, filled circles,
and filled squares, respectively.
B2{\ts}1225+317 (radio-loud) is shown by the filled triangle
because its radio spectrum is unknown.
\label{fig-5}}

\figcaption[Murayama.fig6.ps]{%
Diagram between equivalent width of {\it EW}(C{\sc{} iv}{\ts}$\lambda$1549) and
{\it EW}(Fe{\sc{}~ii}{\ts}$\lambda\lambda$4434--4684).
Large symbols are high-$z$ quasars listed in Table 3 and
small symbols are low-$z$ quasars studied by
Wang {\it et al}.\ (1996b).
Radio-quiet quasars, flat-spectrum radio-loud quasars, and
steep-spectrum radio-loud quasars
are shown by open circles, filled circles,
and filled squares, respectively.
B2{\ts}1225+317 (radio-loud) is shown by the filled triangle
because its radio spectrum is unknown.
\label{fig-6}}

\end{document}